\documentclass[sigconf]{acmart} 

\copyrightyear{2025}
\acmYear{2025}
\setcopyright{acmlicensed}
\acmConference[SIGIR '25] {Proceedings of the 48th International ACM SIGIR Conference on Research and Development in Information Retrieval}{July 13--18, 2025}{Padua, Italy.}
\acmBooktitle{Proceedings of the 48th International ACM SIGIR Conference on Research and Development in Information Retrieval (SIGIR '25), July 13--18, 2025, Padua, Italy}
\acmISBN{979-8-4007-1592-1/25/07}
\acmDOI{10.1145/3726302.3730206}
\AtBeginDocument{%
  }

\usepackage{enumitem}
\usepackage{booktabs}  % 用于美化表格的横线
\usepackage{multirow}  % 用于跨行合并单元格
\usepackage{graphicx}  % 用于缩放表格
\usepackage{amsmath}  % 推荐引入amsmath包以支持更丰富的数学符号
\usepackage{mathrsfs}  % 提供优雅的手写体
\usepackage{CJKutf8} 
\usepackage{subfigure}
\usepackage{}\usepackage{balance} % 导言区引入 

\begin{document}

\title{HeterRec: Heterogeneous Information Transformer for Scalable Sequential Recommendation}


\author{Hao Deng}
\orcid{0009-0002-6335-7405}
\authornote{Contributed equally to this research.} 
\affiliation{%
  \institution{Alibaba International Digital Commerce Group}
   \city{Beijing} 
   \state{} 
   \country{China}
}
\email{denghao.deng@alibaba-inc.com}

\author{Haibo Xing}
\orcid{0009-0006-5786-7627}
\authornotemark[1]
\affiliation{%
  \institution{Alibaba International Digital Commerce Group}
  \city{Hangzhou} 
  \state{} 
  \country{China}
}
\email{xinghaibo.xhb@alibaba-inc.com}

\author{Kanefumi Matsuyama}
\orcid{0009-0002-1365-5375}
\authornotemark[1]
\affiliation{%
  \institution{Alibaba International Digital Commerce Group}
  \city{Hangzhou} 
  \state{} 
  \country{China}
}
\email{kan.matsu@alibaba-inc.com}

\author{Yulei Huang}
\orcid{0009-0004-0378-5330}
\affiliation{
  \institution{Alibaba International Digital Commerce Group}
  \city{Hangzhou} 
  \state{} 
  \country{China}
}
\email{huangyulei.hyl@alibaba-inc.com}

\author{Jinxin Hu}
\orcid{0000-0002-7252-5207}
\authornote{Corresponding author.}
\affiliation{
  \institution{Alibaba International Digital Commerce Group}
  \city{Beijing} 
  \state{} 
  \country{China}
}
\email{jinxin.hjx@lazada.com}

\author{Hong Wen}
\orcid{0009-0006-5786-7627}
\affiliation{
  \institution{Unaffiliated}
  \city{Hangzhou} 
  \state{} 
  \country{China}
}
\email{dreamonewh@gmail.com}

\author{Jia Xu}
\orcid{0000-0003-4061-8262}
\affiliation{%
  \institution{Cyberspace Institute of Advanced Technology, Guangzhou University}
  \city{Guangzhou} 
  \state{} 
  \country{China}
}
\email{xujia@gzhu.edu.cn}

\author{Zulong Chen}
\orcid{0000-0003-0807-6889}
\affiliation{%
  \institution{Alibaba Group}
  \city{Hangzhou} 
  \state{} 
  \country{China}
}
\email{zulong.czl@alibaba-inc.com}

\author{Yu Zhang}
\orcid{0000-0002-6057-7886}
\affiliation{
  \institution{Alibaba International Digital Commerce Group}
  \city{Beijing} 
  \state{} 
  \country{China}
}
\email{daoji@lazada.com}

\author{Xiaoyi Zeng}
\orcid{0000-0002-3742-4910}
\affiliation{
  \institution{Alibaba International Digital Commerce Group}
  \city{Hangzhou} 
  \state{} 
  \country{China}
}
\email{yuanhan@taobao.com}

\author{Jing Zhang}
\orcid{0000-0001-6595-7661}
\affiliation{
  \institution{School of Computer Science, Wuhan University}
  \city{Wuhan} 
  \state{} 
  \country{China}
}
\email{jingzhang.cv@gmail.com}

\renewcommand{\shortauthors}{Hao Deng et al.}
% Sequential Recommendation, Transformer, Heterogeneous Information Extraction
\begin{abstract}
Transformer-based sequential recommendation (TSR) models have shown superior performance in recommendation systems, where the quality of item representations plays a crucial role. Classical representation methods integrate item features using concatenation or neural networks to generate homogeneous representation sequences. While straightforward, these methods overlook the heterogeneity of item features, limiting the transformer's ability to capture fine-grained patterns and restricting scalability. Recent studies have attempted to integrate user-side heterogeneous features into item representation sequences, but item-side heterogeneous features, which are vital for performance, remain excluded. To address these challenges, we propose a Heterogeneous Information Transformer model for Sequential Recommendation (HeterRec), which incorporates Heterogeneous Token Flatten Layer (HTFL) and Hierarchical Causal Transformer Layer (HCT). Our HTFL is a novel item tokenization method that converts items into a heterogeneous token set and organizes these tokens into heterogeneous sequences, effectively enhancing performance gains when scaling up the model. Moreover, HCT introduces token-level and item-level causal transformers to extract fine-grained patterns from the heterogeneous sequences. Additionally, we design a Listwise Multi-step Prediction (LMP) Loss function to further improve performance. Extensive experiments on both offline and online datasets show that the HeterRec model achieves superior performance.
\end{abstract}

\begin{CCSXML}
<ccs2012>
   <concept>
       <concept_id>10002951.10003317.10003338.10003342</concept_id>
       <concept_desc>Information systems~Similarity measures</concept_desc>
       <concept_significance>500</concept_significance>
       </concept>
 </ccs2012>
\end{CCSXML}
\ccsdesc[500]{Information systems~Recommender systems}
\keywords{Sequential Recommendation, Heterogeneous Features, Scaling Up}
\maketitle
\section{Introduction}
\begin{figure}[htbp]
\setlength{\abovecaptionskip}{0.cm}
 \includegraphics[width=0.4\textwidth]{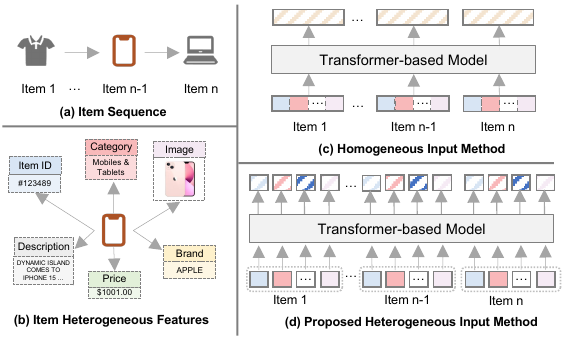}
  \caption{Comparison of TSR input methods: (c) Traditional concatenation-based homogeneous item sequences vs. (d) The proposed heterogeneous token sequences. }
  \Description{xx.}
  \label{fig:intro_seq}
\vspace{-0.65cm} %调整图片与上文的垂直距离
\end{figure}
% Sequential recommendation models aim to predict the next item a user is likely to engage with based on their historical behaviors. In this context, the user's behavior sequence serves as a "corpus" for modeling their interests, with each individual behavior functioning as a "word" within this corpus \cite{kang2018self, liao2023llara}. Traditionally, sequential recommendation models have relied on Recurrent Neural Networks (RNNs) to capture implicit patterns in user behaviors \cite{hidasi2015session, zhao2018go}. However, in recent years, Transformer-based Sequential Recommendation (TSR) models have exhibited superior performance and have been widely adopted in industrial recommendation systems \cite{zhai2024actions, pancha2022pinnerformer, sun2019bert4rec, fan2022modeling}. Specifically, TSR models harness the transformer's advanced information extraction capabilities by first encoding user behaviors into a dense sequence of item representations. These representations are then processed by the transformer to generate comprehensive user interest representations.
Sequential recommendation models predict the next item a user might be interested in, based on their past behaviors. The behavior sequence functions as the "corpus" for modeling user interests, with each behavior acting as a "word" in the corpus \cite{kang2018self, liao2023llara}. Traditional sequential recommendation models primarily use Recurrent Neural Networks (RNNs) to capture implicit patterns within behaviors \cite{hidasi2015session,zhao2018go}. In recent years, Transformer-based Sequential Recommendation (TSR) models have demonstrated superior performance and are now widely used in industrial recommendation systems \cite{zhai2024actions,pancha2022pinnerformer,sun2019bert4rec}. Specifically, to fully leverage the transformer's information extraction capabilities, TSR models first encode user behaviors into a dense sequence of item representations, which are then processed by the transformer to generate user interest representations.

\textbf{Related Works \& Challenges.} The volume of input information determines the upper limit of a model's capacity \cite{yang2022does}, making the enhancement of item representations within sequences a key research focus for TSR models in recent years. Existing approaches \cite{chen2022intent, li2022coarse} commonly extract heterogeneous item features, such as text, images, and categories, to enrich item information, as illustrated in Figure \ref{fig:intro_seq} (b). To incorporate such heterogeneous information into TSR models, traditional methods \cite{kang2018self, sun2019bert4rec, pancha2022pinnerformer} typically concatenate all item features or use neural networks to aggregate them, producing a homogeneous item representation sequence, as depicted in Figure \ref{fig:intro_seq} (c). However, these straightforward aggregation techniques limit the transformer's ability to capture \textbf{fine-grained heterogeneous patterns} and hinder the scalability of TSR models. Drawing inspiration from large language models (LLMs) \cite{kaplan2020scaling}, TSR models are expected to achieve significant performance gains by increasing the number of parameters (\textit{e.g.}, stacking more layers). Nonetheless, existing research has largely overlooked the design of input methods for TSR, limiting their ability to fully achieve \textbf{performance gains when scaling up} the TSR model. Specifically, while some studies \cite{zhai2024actions, zhang2019feature} have explored converting TSR input sequences from homogeneous to heterogeneous representations, many crucial heterogeneous item features (\textit{e.g.}, text, images, and categories) remain underused or excluded from the heterogeneous representation sequence. For instance, HSTU \cite{zhai2024actions} incorporates heterogeneous user-side information, such as followed authors and user profiles, into TSR input sequences. The absence of these heterogeneous item features limits the model's ability to capture fine-grained implicit relationships between items, which are vital for improving performance. Therefore, scaling up TSR models without addressing input design challenges may lead to suboptimal performance. 
% This research gap highlights the need for novel methods to incorporate heterogeneous features into TSR input sequences, potentially leading to more accurate and personalized recommendations, especially in large-scale recommendation systems.
 
 To address these challenges, we propose a Heterogeneous Information Transformer model for Sequential Recommendation (HeterRec), inspired by tokenization techniques in LLMs. HeterRec significantly enhances performance and achieves performance gains when scaling up the TSR models. First, HeterRec introduces a novel item tokenization method in the Heterogeneous Token Flatten Layer (HTFL), treating each item as a "word" and tokenizing it into a fine-grained heterogeneous token (feature) set. Then, as shown in Figure \ref{fig:intro_seq} (d), we flatten and organize the token set into a heterogeneous representation sequence based on the order of user behavior. Our item tokenization method enables the Hierarchical Causal Transformer Layer (HCT) to fully leverage the transformer's information extraction capabilities and improve scalability. Furthermore, to assist HeterRec in capturing fine-grained patterns, we design a Listwise Multi-step Prediction (LMP) loss function, incorporating token-level learning tasks alongside item-level tasks. Our key contributions are summarized below:	

\begin{itemize}[noitemsep, topsep=0pt, leftmargin=*]
% \item We introduce the HeterRec model, which incorporates a novel item tokenization method. HeterRec generates a heterogeneous representation sequence with critical item features, enhancing the transformer's ability to capture fine-grained implicit patterns and scalability.
\item We introduce the HeterRec model, which incorporates a novel item tokenization method to generate a heterogeneous representation sequence with critical item features and employs HCT to enhance the transformer's ability to capture fine-grained implicit patterns and improve scalability.	
\item We integrate a novel LMP loss function that includes both token-level and item-level tasks to help HeterRec learn fine-grained patterns.
\item Extensive experiments on two datasets validate the superiority of HeterRec, which has been deployed in an online system.
\end{itemize}
\section{OUR APPROACH}
% \section{Problem Formulation}
\subsection{Problem Statement}
For a given user $u$, the chronological sequence of interactions is represented as $\mathcal{S}=\{i_{1},i_{2},...i_{t},...i_{T}\}$, where $T$ is the maximum length of $\mathcal{S}$, and $i_{t} \in I$ denotes the item interacted with by the user at the $t$-th step. The embeddings set of heterogeneous features for an item $i$ is defined as $\textbf{d}_{i}=\{ \textbf{e}^{1}_{i}, \textbf{e}^{2}_{i}, ..., \textbf{e}^{K}_{i}\}$, where $K$ is the maximum length of $\textbf{d}_{i}$. Thus, the item sequence of a user $u$ can be further expressed as: $S=\{ ( i_{1}, \textbf{d}_{1} ), ( i_{2},\textbf{d}_{2}), ..., ( i_{T},\textbf{d}_{T}) \}$. The goal of TSR is to predict the next most likely item $i_{T+1} \in  I$ that the user will interact with at time $T+1$, given the user’s behavior sequence $\mathcal{S}$. Formally, this can be expressed as $p(i_{T+1}|S)$. Following prior works \cite{pancha2022pinnerformer, zhai2024actions}, we employ the TSR model to the recall stage of a recommendation system. A typical two-tower embedding-based retrieval (EBR) framework uses $\mathcal{S}$ and $\textbf{d}_{T+1}$ as inputs to generate user and item vectors, respectively \cite{huang2020embedding}. During training, EBR employs contrastive learning to differentiate positive items from negative items \cite{mikolov2013distributed}.
% \section{OUR APPROACH: HeterRec}
\begin{figure}[htbp]  
  \centering 
  \includegraphics[width=0.47\textwidth]{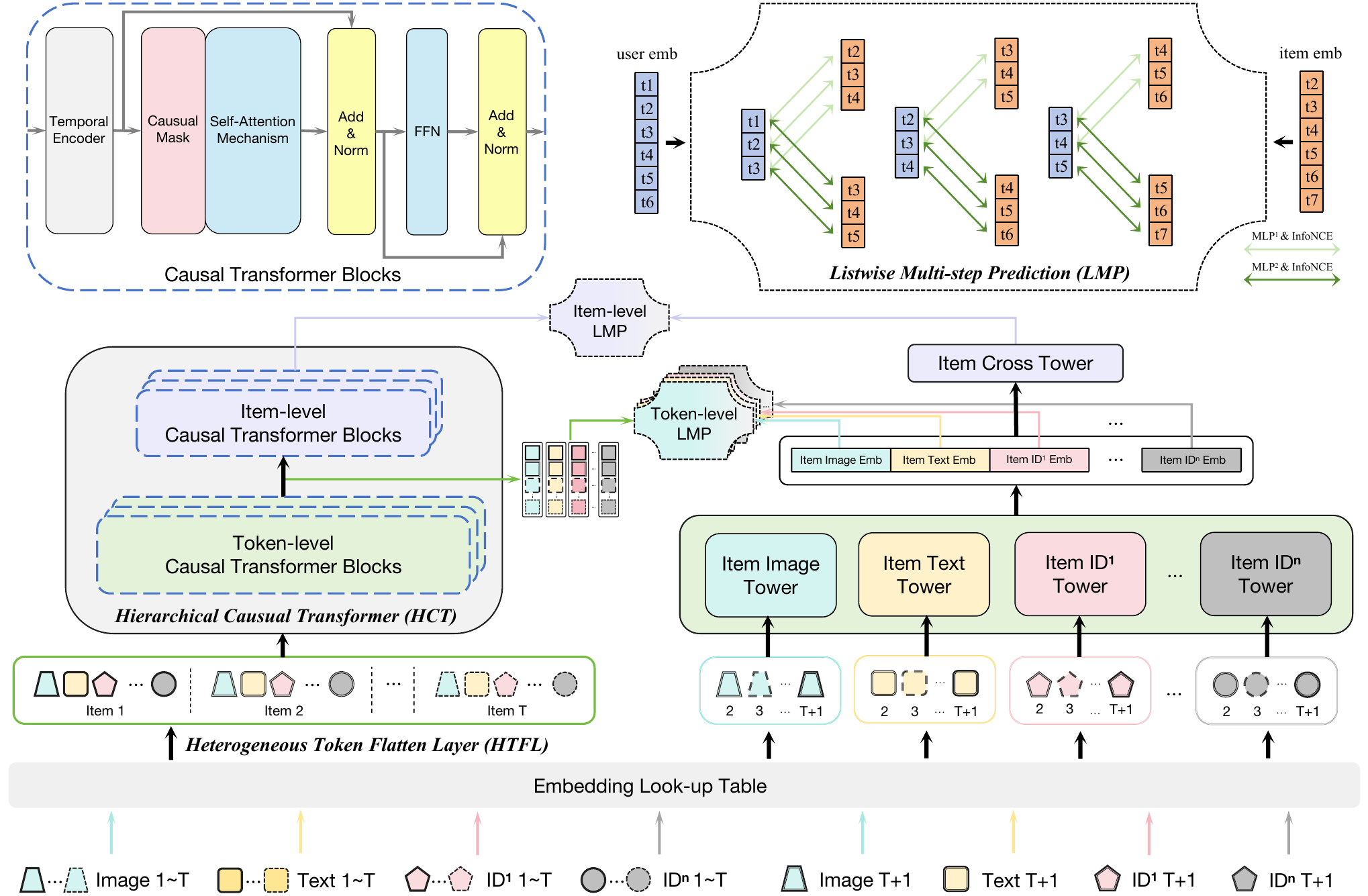}  % 调整宽度为单栏比例
  \caption{Our proposed HeteroRec framework.}
  \Description{xx.}
  \label{fig:token_framework_v1}
  \vspace{-0.7cm} % 调整图片与上文的垂直距离
\end{figure}
%%We assume that the length of the preceding neighbor window $r_t=1$, the number of preceding tokens selected by each expert $r_m=1$, the number of experts $K=3$ and the minimum number of valid votes $N_v=2$.
\subsection{Overwiew For HeterRec}
Figure \ref{fig:token_framework_v1} illustrates the HeterRec framework. The highlight of our work lies in the user-side modeling in the EBR model, which comprises two key components: \textbf{(a) Heterogeneous Token Flatten Layer (HTFL)}: This layer is designed to flatten heterogeneous item features (\textit{e.g.}, visual, textual, categorical features) along the chronological order of behavior. We categorize these heterogeneous features into three types and design separate embedding encoding strategies to map each type into continuous numerical spaces. \textbf{(b) Hierarchical Causal Transformer (HCT)}: This module includes token-level causal transformer blocks to extract fine-grained patterns from the heterogeneous sequence, followed by item-level causal transformer blocks to integrate user-side information into the final embedding. For the item-side modeling, we use hierarchical Multi-Layer Perceptrons (MLPs) to first generate token-level embeddings and subsequently produce the final item embedding. The framework is ultimately supervised by our proposed \textbf{Listwise Multi-step Prediction (LMP)}.
\subsection{Heterogeneous Token Flatten Layer}
This section introduces the item tokenization method. To generate heterogeneous token sequences for TSR, we ensure that the embedding dimensions of all heterogeneous features are consistent. Therefore, we categorize the common features in recommendation systems into three categories: \textit{Categorical Features} (\textit{e.g.}, item IDs and categories), \textit{Numerical Features} (\textit{e.g.}, item price) and \textit{Multimodal Features} (\textit{e.g.}, image and text). Then, we employ three distinct encoding techniques to map these features into continuous numerical spaces with consistent embedding dimensions.
% In this section, we introduce the token-level flatten strategy, which actually relies on the length-consistency among all input features. Therefore, we categorize the common features in recommendation systems into three main categories: \textit{Categorical Features} (\textit{e.g.}, item IDs, categories and brands), \textit{Numerical Features} (\textit{e.g.}, item price and posterior click-through rate) and \textit{Multimodal Features} (\textit{e.g.}, image and text). Then, we utilize three distinct embedding encoding techniques to map these features into the length-consistency continuous numerical spaces. 

\textit{Categorical Features} can be directly transformed into the numerical space using an embedding lookup matrix $\textbf{E}_{id_j} \in \mathbb{R}^{N_{id_j} \times d_f}$, where $N_{id_j}$  denotes the number of unique values of the $j$-th categorical feature and $d_f$ is the embedding dimension. \textit{Numerical Features} are first discretized into several bins using the binning methods \cite{guo2021embedding,malak2023distributed}. Then, the $j$-th numerical feature can be transformed into the numerical space using an embedding lookup matrix $\textbf{E}_{nb_j} \in \mathbb{R}^{N_{nb_j} \times d_f}$, where $N_{nb_j}$  denotes the number of bins for the $j$-th numerical feature. As for \textit{Multimodal Features}, training a large multimodal model for an advertising recommendation system is \textbf{extremely expensive} and often suffers from \textbf{poor transferability}. Therefore, we leverage external, pre-trained multimodal models to extract the visual and textual representations of items. However, directly using pre-trained representations can lead to suboptimal performance due to discrepancies between the pre-training objectives and downstream tasks \cite{dodge2020fine}. To address this issue, we quantize the multimodal representations into cluster IDs and subsequently map them to trainable embeddings. Consider the visual representation $\textbf{v}\in \mathbb{R}^{N_{item} \times d_{v}}$ as an example, where $N_{item}$ denotes the number of unique items. We evenly divide $\textbf{v}$ into $Z$ groups, denoted as $\{\textbf{v}^1, \textbf{v}^2, ..., \textbf{v}^Z \}$. In each group, we partition the values into $q$ quantiles and encode these quantiles into unique tokens $\{\textbf{t}_v^1, \textbf{t}_v^2, ..., \textbf{t}_v^Z \}$, where the $k$-th token is represented as $\textbf{t}_v^k \in \mathbb{R}^{q^{{d_v}/Z}}$. Subsequently, we apply an embedding lookup matrix $\textbf{E}_{v} \in \mathbb{R}^{q^{{d_v}/Z} \times d_f/z}$ to each token to map it into a numerical space. Finally, we concatenate these tokens into a single representation and obtain a numerical space with $d_f$ dimensions.

Then we divide the total sequence into multiple tokens, which will be subsequently used as input representations for the HCT. The heterogeneous information sequence can be represented as $\textbf{H}_{0} =\{ \textbf{e}^{1}_{1}, \textbf{e}^{2}_{1}, ...,\textbf{e}^{K}_{1}, \textbf{e}^{1}_{2}, \textbf{e}^{2}_{2}, ...,\textbf{e}^{K}_{2},  ..., \textbf{e}^{1}_{T}, \textbf{e}^{2}_{T}, ...,\textbf{e}^{K}_{T} \}$, where $K$ denotes the total number of the three types of features described above.
% \begin{equation}
% \begin{aligned}
% \begin{split}
% \bar{\textbf{H}}_{0} =\{ \textbf{h}^{1}_{1}, \textbf{h}^{1}_{2}, ...,\textbf{h}^{1}_{T}, \textbf{h}^{2}_{1}, \textbf{h}^{2}_{2}, ...,\textbf{h}^{2}_{T},  ..., \textbf{h}^{K}_{1}, \textbf{h}^{K}_{2}, ...,\textbf{h}^{K}_{T} \}.
% \end{split}
% \end{aligned}
% \label{Eq.2}
% \end{equation}
\subsection{Hierarchical Causal Transformer}
\label{sec:MMFT}
%which consists of two transformer-based submodules: \textit{token-level Causal Transformer Blocks} and \textit{item-level Causal Transformer Blocks}. 
In this section, we introduce the backbone of our HeterRec model, which consists of two transformer-based submodules. First, $\textbf{H}_0$ is fed into a stack of $N_1$ \textit{token-level Causal Transformer Blocks} to capture fine-grained patterns over the flattened heterogeneous sequence. The fine-grained patterns can be categorized into three types as follows: (1) interaction signals of heterogeneous features within the same item, (2) interaction signals of homogeneous features across different items, and (3) interaction signals of heterogeneous features across different items. In this way, we obtain the heterogeneous representation $\textbf{H}_{N_1}$. After extracting multi-patterns at the token level, we need to integrate the user's interest vector at a higher dimension. Therefore, an MLP layer is applied to fuse the tokens within each item. The output of this layer $\bar{\textbf{H}}_0$ is then fed into a stack of $N_2$ \textit{item-level Causal Transformer Blocks} to obtain the final user embedding $\bar{\textbf{H}}_{N_2} \in \mathbb{R}^{d_k \times T}$. 

The basic structure of the two submodules is similar (both are Causal Transformer Blocks), but they differ in some components. Therefore, we will first introduce the Causal Transformer Blocks (shown in the top-left corner of Figure \ref{fig:token_framework_v1}) and then highlight the differences between the two submodules. The Temporal Encoder and Self-Attention Mechanism with a Causal Mask are defined as:
\begin{equation}
\begin{aligned}
\text{Attention}(Q, K, V) = \text{softmax}\left(\frac{QK^T+P^{tg}}{\sqrt{d_k}} + M\right)V,
\end{aligned}
\label{Eq.self_atten1}
\end{equation}
\begin{equation}
\begin{aligned}
P^{tg}_{i,j}=LookUp(TGBkt({ts}_j-{ts}_i), \textbf{E}_{tg}), 1 \le i \le j \le T,
\end{aligned}
\label{Eq.TimeEmbTable}
\end{equation}
\begin{equation}
\begin{aligned}
M_{i,j} = 
\begin{cases} 
0, & \text{if } i \geq j \\
-\infty, & \text{if } i < j
\end{cases},
\end{aligned}
\label{Eq.M_mask}
\end{equation}
where \( Q \) is the query matrix, \( K \) is the key matrix, \( V \) is the value matrix, \( d_k \) is the dimension of the key vectors, and \( M \) is the causal mask matrix. Classical TSR models fail to capture temporal changes, which may limit the improvement of model performance. To address this issue, we incorporate a learnable time gap representation matrix (denoted as $P_{tg} \in \mathbb{R}^{T \times T}$) into the relevance matrix (denoted as $QK^T$). ${ts}_i$ denotes the timestamp at which the $i$-th user behavior occurs. $TGBkt(\cdot)$ is used to bucketize the generated time gap feature. After that, we can obtain the dense representation of the time gap in the user sequence by looking up an embedding table (denoted as $E_{tg}$). The causal mask \( M \) is an upper triangular matrix used to mask future tokens. 

It is worth noting that tokens of the same item share the same time gap representation. Similarly, The mask \( M \) will mask all tokens from the current position to future items while preserving all past tokens. The multi-head self-attention mechanism is defined as:
\begin{equation}
\begin{aligned}
\text{MultiHead}(Q, K, V) = \text{Concat}(\text{head}_1, \text{head}_2, \ldots, \text{head}_h) W^O,
\end{aligned}
\label{Eq.MultiH1}
\end{equation}
\begin{equation}
\begin{aligned}
 \text{head}_i = \text{Attention}(XW_i^Q, XW_i^K, XW_i^V),
\end{aligned}
\label{Eq.MultiH2}
\end{equation}
where \( W_i^Q, W_i^K, W_i^V \) are weight matrices, \( X \) is the input representation, \( W^O \) is the output linear transformation matrix, and \( h \) is the number of heads. The feed-forward neural network is defined as:
\begin{equation}
\begin{aligned}
\text{FFN}(x) = \max(0, xW_1 + b_1)W_2 + b_2,
\end{aligned}
\label{Eq.FFN1}
\end{equation}
where \( W_1 \) and \( W_2 \) are weight matrices, \( b_1 \) and \( b_2 \) are bias terms and \( \max(0, \cdot) \) is the ReLU activation function. A complete Causal Transformer Block includes multi-head self-attention and a feed-forward neural network, with residual connections and layer normalization:
\begin{equation}
\begin{aligned}
\text{CausalTransformer}(x) = \text{LayerNorm}(x + \text{Sublayer}(x)),
\end{aligned}
\label{Eq.LN2}
\end{equation}
where \( \text{Sublayer}(x) \) can be either the multi-head self-attention or the feed-forward neural network , \( \text{LayerNorm} \) is layer normalization.

\begin{table*}[htbp!]
\centering
    \caption{Performance across two datasets is evaluated using Recall@N and nDCG@N metrics. The last row shows the relative improvement of our method over the best baseline.}
    \centering
    \vspace{-12pt}  % 减少标题和表格之间的间距
    \resizebox{0.9\textwidth}{!}{  % 缩放表格以适应双栏中的一栏宽度
        \begin{tabular}{c|ccc|ccc|ccc|ccc}
            \toprule
            \multirow{2}{*}{\textbf{Method}}  & \multicolumn{6}{c|}{\textbf{Industrial Dataset}} & \multicolumn{6}{c}{\textbf{Amazon Elecs Dataset}} \\
            \cline{2-13}
            & Recall@50 & Recall@100 & Recall@200 &  nDCG@50  &  nDCG@100  &  nDCG@200 & Recall@50 & Recall@100 & Recall@200 &  nDCG@50  &  nDCG@100  &  nDCG@200  \\
            \midrule
            SASRec & 12.08\% & 14.58\% & 17.40\% & 8.17\% & 9.04\% & 10.90\%  & 15.12\% & 25.86\% & 38.04\% & 11.23\% & 15.36\% & 19.76\% \\
            BERT4Rec & 11.53\% & 14.71\% & 18.30\% & 8.75\% & 9.86\% & 10.96\% & 16.88\% & 27.50\% & 41.98\% & 14.12\% & 17.98\% & 23.48\% \\
            % PinnerFormer & 12.10\% & 15.13\% & 18.65\% & 9.43\% & 10.48\% & 11.56\% & \underline{19.65}\% & 28.43\% & 43.48\% & \underline{14.72}\% & 18.34\% & 24.20\% \\
            % STRec & 12.57\% & 15.82\% & 19.42\% & 9.63\% & 10.76\% & 11.87\% & 18.88\% & 29.50\% & 43.04\% & 14.44\% & 18.50\% & 23.98\% \\
            PinnerFormer & 12.57\% & 15.82\% & 19.42\% & 9.63\% & 10.76\% & 11.87\% & \underline{19.65}\% & 29.50\% & 43.04\% & \underline{14.72}\% & 18.50\% & 23.98\% \\
            HSTU & \underline{12.83}\% & \underline{16.17}\% & \underline{19.70}\% & \underline{10.13}\% & \underline{10.99}\% & \underline{12.14}\% & 19.56\% & \underline{29.82}\% & \underline{43.60}\% & 14.64\% & \underline{18.96}\% & \underline{24.27}\% \\
            \midrule
            HeterRec (Ours) & \textbf{13.20}\% & \textbf{16.78}\% & \textbf{20.05}\% & \textbf{10.57}\% & \textbf{11.54}\% & \textbf{12.54}\% & \textbf{20.40}\% & \textbf{31.06}\% & \textbf{45.14}\% & \textbf{15.10}\% & \textbf{20.02}\% & \textbf{25.35}\%  \\
            \midrule
            Improvement & 2.84\% & 3.81\% & 1.76\% & 4.33\% & 5.00\% & 3.30\% & 4.29\% & 4.17\% & 3.52\% & 2.58\% & 5.59\% & 4.47\% \\
            \bottomrule
            
        \end{tabular}
    }
    \label{table:results}
    \vspace{-0.33cm} 
\end{table*}
% \vspace{-0.4cm} 

\subsection{Listwise Multi-step Prediction}
\label{sec:LMP}
Recent research on LLMs \cite{gloeckle2024betterfasterlarge, deepseekai2024deepseekv3technicalreport} show that the Multi-step Token Prediction (MTP) task enhances long-term token dependencies while mitigating sequence fluctuations. Inspired by this, we propose the LMP task. Firstly, the user embedding output can be redefined as $U=\bar{\textbf{H}}_{N_2}$. We also express $U$ as $\{U_{1}, U_{2}, ... , U_{T}\}$, where $U_{t} \in \mathbb{R}^{d_k}$ is the $t$-th item of the user sequence. Similarly, the output embedding of the final item encoder is defined as $V = \{V_{2}, V_{3}, ... , V_{T+1}\}$, where $V_{t} \in \mathbb{R}^{d_k}$. We use in-batch negative sampling to generate negative samples, with the positive and negative \textit{logits} defined as follows:
\begin{equation}
\begin{aligned}
\begin{split}
Lgt^{{j},t}_{{i},pos} = {{MLP}_i(U_{t}^{j})}^\top V_{t+i}^{j},
\end{split}
\end{aligned}
\label{Eq.cy.logit1}
\end{equation}
\begin{equation}
\begin{aligned}
\begin{split}
Lgt^{j,k,t}_{i,neg} = {{MLP}_i(U_{t}^{j})}^\top V_{t+i}^{k}, j \neq k,
\end{split}
\end{aligned}
\label{Eq.cy.logit2}
\end{equation}
where $i$ denotes the $i$-th step token prediction, $j$ and $k$ are sample indices in a mini-batch, and ${MLP}_i(\cdot)$ is the linear transformation of the user embedding at the $i$-th step. The objective function used here is the InfoNCE loss \cite{mikolov2013distributed} defined as follows:
{\footnotesize\begin{equation}
\begin{aligned}
\begin{split}
L_{step_1}^{item}=\sum\limits_{j \in B, t \in T} \frac{e^{Lgt^{{j},t}_{1,pos}/{\tau}}}{1+e^{Lgt^{{j},t}_{1,pos}/{\tau}}+\sum\limits_{k \in B,k \neq j} e^{Lgt^{j,k,t}_{1,neg}/{\tau}}},
\end{split}
\end{aligned}
\label{Eq.cy.mtploss1}
\end{equation}
}
{\footnotesize\begin{equation}
\begin{aligned}
\begin{split}
\small L_{step_i}^{item}=\sum\limits_{j \in B, t \in T} \frac{\mathbb{I}(e^{Lgt^{{j},t}_{i-1,pos}/{\tau}}>e^{Lgt^{{j},t}_{i,pos}/{\tau}}+\lambda_{m})*e^{Lgt^{{j},t}_{i,pos}/{\tau}}}{1+e^{Lgt^{{j},t}_{i,pos}/{\tau}}+\sum\limits_{k \in B,k \neq j} e^{Lgt^{j,k,t}_{i,neg}/{\tau}}}, i > 1,
\end{split}
\end{aligned}
\label{Eq.cy.mtploss2}
\end{equation}}
where $\mathbb{I}(\cdot)$ is the characteristic function,  $\tau$ is the temperature hyperparameter and $\lambda_m$ is the margin hyperparameter that controls the influence of multi-step logits. Additionally,  we construct the \textbf{token-level LMP} task to enhance the heterogeneous representation. The final loss function is defined as follows:
\begin{equation}
\begin{aligned}
\begin{split}
L_{total}=\sum\limits_{i} ({L_{step_i}^{item}}+\sum\limits_{s} {L_{step_i}^{token_s}}).
\end{split}
\end{aligned}
\label{Eq.cy.mtploss2}
\end{equation}
\section{Experiments}
\subsection{Experimental Setup}
\textbf{Datasets \& Evaluate Settings.} We conduct experiments on \textbf{two datasets}: the Amazon dataset \cite{he2016ups} and a large industrial dataset from the advertising systems of a leading Southeast Asian e-commerce platform. The Amazon dataset uses the Electronics subset, which contains extensive user reviews and comprehensive metadata like product titles and categories. Textual features are extracted using Sentence-Transformers \citep{reimers2019sentence} following \citep{zhou2023bootstrap}, while visual features are sourced from \citep{ni2019justifying}. For the industrial dataset, we follow HSTU \cite{zhai2024actions}, using training data with three user behaviors (click, add-to-cart, and conversion) from the past 180 days. For HeterRec, we set the maximum sequence length $T$ to 256, both $N_1$ and $N_2$ in HCT to 2, $\lambda_m$ to 1, and $N_{step}$ to 3.	
\\
% \textbf{Baselines \& Metric.} We evaluate the HeterRec by comparing it with \textbf{five models}: SASRec \cite{kang2018self}, BERT4Rec \cite{sun2019bert4rec}, PinnerFormer \cite{pancha2022pinnerformer}, STRec \cite{li2023strec} and  HSTU \cite{zhai2024actions}. Offline metrics include Recall@N and nDCG@N ($N \in \{50, 100, 200\}$), as detailed in \cite{zheng2022multi}.
\textbf{Baselines \& Metrics.} We evaluate HeterRec by comparing it with \textbf{four representative methods}: SASRec \cite{kang2018self}, BERT4Rec \cite{sun2019bert4rec}, PinnerFormer \cite{pancha2022pinnerformer}, and  HSTU \cite{zhai2024actions}. Offline metrics include Recall@N and nDCG@N ($N \in \{50, 100, 200\}$), as detailed in \cite{zheng2022multi}.
\vspace{-7pt}
\subsection{Experimental Results}
\textbf{Overall Performance.} The comparative results in Table \ref{table:results} lead to the following observations: (1) Compared to SASRec, transformer-based models significantly improve user behavior modeling. (2) Compared to BERT4Rec, PinnerFormer demonstrates that the MTP task outperforms the Masked Language Model task \cite{kenton2019bert} in TSR. (3) HeterRec outperforms PinnerFormer on recall@N and nDCG@N metrics, thanks to the LMP loss function, which incorporates fine-grained token-level learning tasks for enhanced performance. (4) HeterRec outperforms HSTU on recall@N and nDCG@N metrics, thanks to HTFL, which leverages diverse item features and transforms homogeneous sequences into heterogeneous ones through item tokenization method. Additionally, the token-level and item-level causal transformer blocks in HCT maximize the transformer's pattern extraction ability, boosting the performance of TSR models.	
\begin{figure}
  \includegraphics[width=0.4\textwidth]{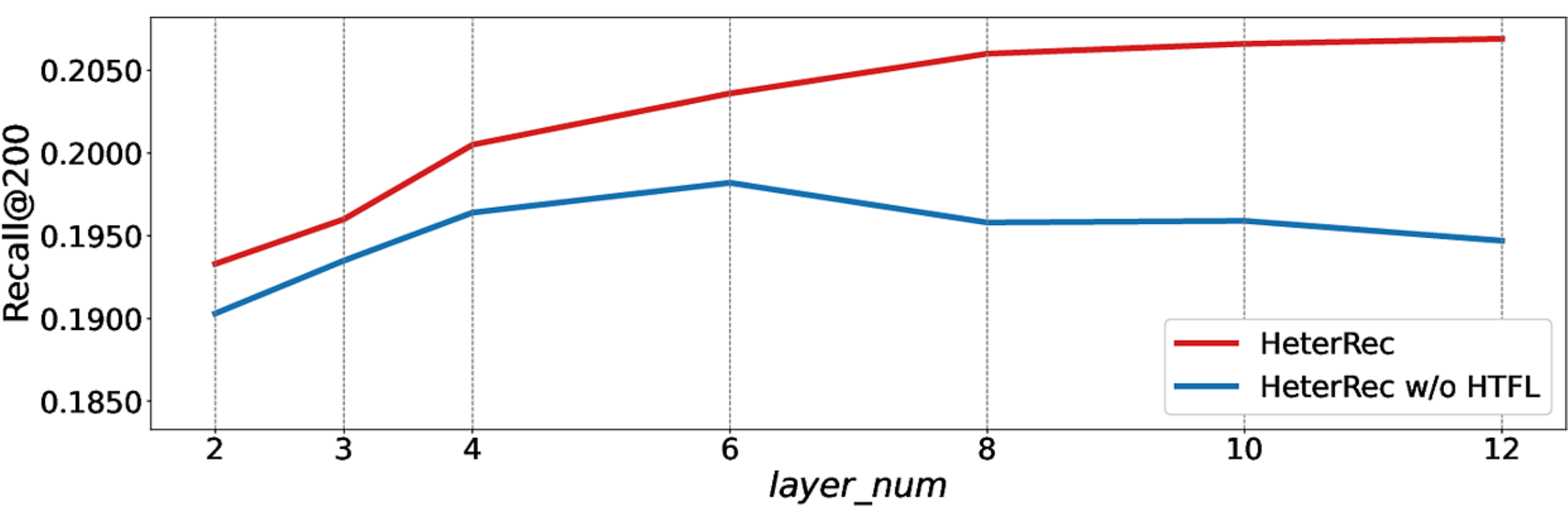}
  \vspace{-8pt}
  \caption{The scalability of HeterRec, with and without HTFL, across different numbers of layers.}
  \Description{xx.}
  \label{fig:scale_up01}
  % \vspace{-15pt}
\end{figure}
\\
\textbf{Scaling Up for TSR Model.} To assess scalability, we evaluated HeterRec's performance and the impact of HTFL across different numbers of layers, as shown in Figure \ref{fig:scale_up01}. We tested seven configurations, $(N_1, N_2) \in \{ (1,1),(1,2),(2,2),(3,3),(4,4),(5,5),(6,6) \}$, to compare the scalability of HeterRec with and without HTFL. The results revealed two key observations: (1) HeterRec achieves notable performance gains as the number of transformer layers increases. (2) HTFL greatly boosts HeterRec's performance during TSR scaling. These improvements stem from the novel item tokenization method and heterogeneous representation sequences, which fully leverage the transformer architecture's potential.
% \vspace{-14pt}
\begin{table}[htbp!]
\caption{Ablation Study on the Industrial Dataset.}
\vspace{-12pt}  % 减少标题和表格之间的间距
\centering
\resizebox{0.42\textwidth}{!} {
\begin{tabular}{l|cccccc}
\toprule
\multirow{1}{*}{Method} & \multicolumn{1}{c}{\centering Recall@50} & \multicolumn{1}{c}{\centering Recall@100} & \multicolumn{1}{c}{\centering Recall@200} & \multicolumn{1}{|c}{\centering nDCG@50} & \multicolumn{1}{c}{\centering nDCG@100} & \multicolumn{1}{c}{\centering nDCG@200} \\ 
 \midrule
\textbf{HeterRec}
&\textbf{0.1320}&\textbf{0.1678}&\textbf{0.2005}&\multicolumn{1}{|c}{\textbf{0.1057}}&\textbf{0.1154}&\textbf{0.1254}\\ 
w/o HTFL &0.1265&0.1601&0.1964&\multicolumn{1}{|c}{0.1015}&0.1108&0.1215\\
\small {\hspace{5pt} ---w/o MFK} &\small {0.1309}&\small {0.1664}&\small {0.1989}&\multicolumn{1}{|c}{\small {0.1050}}&\small {0.1147}&\small {0.1247}\\ 
w/o HCT &0.1285&0.1652&0.1978&\multicolumn{1}{|c}{0.1038}&0.1129&0.1221\\ 
w/o LMP &0.1293&0.1658&0.1975&\multicolumn{1}{|c}{0.1042}&0.1134&0.1229\\ 
w/o t-LMP &0.1310&0.1665&0.1993&\multicolumn{1}{|c}{0.1044}&0.1136&0.1243\\ 
% w/o HTFL &0.1245&0.1601&0.1915&\multicolumn{1}{|c}{0.1015}&0.1108&0.1192\\
% \small {\hspace{5pt} ---w/o QMM} &\small {0.1299}&\small {0.1654}&\small {0.1979}&\multicolumn{1}{|c}{\small {0.1050}}&\small {0.1147}&\small {0.1247}\\ 
% w/o HCT &0.1265&0.1612&0.1928&\multicolumn{1}{|c}{0.1022}&0.1114&0.1211\\ 
% w/o LMP &0.1273&0.1622&0.1935&\multicolumn{1}{|c}{0.1033}&0.1127&0.1224\\ 
% w/o t-LMP &0.1260&0.1635&0.1983&\multicolumn{1}{|c}{0.0984}&0.1086&0.1193\\ 
\bottomrule
\end{tabular}
}
\label{table:ablation}
\end{table}
% \vspace{-10pt}
\\
\textbf{Ablation Study.} We further explore five variants of HeterRec: (1) without (w/o) HTFL (replaced by the concat-based method), (2) w/o MFK (removing multimodal feature tokenization and using embeddings from the pre-trained model), (3) w/o HCT (replaced by a standard transformer), (4) w/o LMP (replaced by MTP), and (5) w/o token-level LMP. From Table \ref{table:ablation}, we draw the following insights: (1) HTFL significantly boosts the TSR model's ability to capture user interests. Additionally, tokenizing multimodal embeddings reduces inconsistencies between TSR and pre-trained tasks, while offering more learnable heterogeneous tokens for HTFL. (2) Compared to MTP, LMP greatly improves HeterRec's performance, and token-level LMP further enhances the learning of finer-grained patterns.
% \setlength{\textfloatsep}{10pt} % 设置浮动体与上下文字的间距为 10pt
% \subsection{Ablation Study}
% \subsection{Hyperparameters Sensitivity Analysis}
% \textbf{Hyperparameters Sensitivity Analysis.} We conducted experiments to identify the optimal hyperparameters, as illustrated in Figure \ref{fig:token_sensitive_v1}. In the MEHF layer, both $r_t$ for PNW and $r_m$ for MSPT play a crucial role in enhancing HeterRec's efficiency, controlling uncertainty within behavior sequences. Furthermore, due to the noise in behaviors, an excessively large $r_t$ in LMP significantly impairs performance, while the model stabilizes once $N_{step} > 3$.
\\
\textbf{Online Experiments.} We conducted an online A/B test on an advertising system from January 1 to 10, 2025. The HeterRec model demonstrated a consistent increase of \textbf{1.93\%} in \textbf{Advertising Revenue} and \textbf{1.63\%} in \textbf{Click-Through Conversion Rate (CTCVR)}, demonstrating a significant improvement in industrial scenarios. These results further highlight the effectiveness of HeterRec.
\vspace{-8pt}
\section{Conclusion}
In this paper, we propose a Heterogeneous Information Transformer model (HeterRec) to address the limitations of existing Transformer-based Sequential Recommendation (TSR) models. HeterRec introduces a novel item tokenization method that treats each item as a "word" and tokenizes it into a fine-grained heterogeneous token (feature) set, generating a heterogeneous representation sequence. It further integrates a Hierarchical Causal Transformer to fully exploit the transformer's ability to capture fine-grained patterns and enhance scalability. Importantly, HeterRec achieves significant performance gains when scaling up the TSR model. Extensive offline experiments on real-world datasets, along with online deployment in an advertising system, validate HeterRec's superior performance and practical value. In summary, HeterRec presents a novel, scalable solution for TSR models, providing valuable insights for future research in this domain.
\bibliographystyle{ACM-Reference-Format}
\balance
\bibliography{arxiv}

% \appendix









\end{document}